\documentclass[PRB,showpacs,superscriptaddress,amsmath,amsfonts,amssymb]{revtex4-1}
\usepackage{natbib}
\usepackage[T1]{fontenc}
\usepackage[latin9]{inputenc}
\usepackage{units}
\usepackage{amssymb}
\usepackage{graphicx}
\usepackage{wasysym}
\usepackage{grffile} 
\usepackage{color}
\usepackage{amsmath}
\usepackage{siunitx}

\begin{document}

\title{
Early-stage dynamics of metallic droplets embedded in the nanotextured Mott insulating phase of V$_2$O$_3$}

\author{A. Ronchi}
\email{andrea.ronchi@unicatt.it}
\affiliation{Department of Physics and Astronomy, KU Leuven, Celestijnenlaan 200D, 3001 Leuven, Belgium}
\affiliation{Department of Physics, Universit\`a Cattolica del Sacro Cuore, Brescia I-25121, Italy}
\affiliation{ILAMP (Interdisciplinary Laboratories for Advanced Materials Physics), Universit\`a Cattolica del Sacro Cuore, Brescia I-25121, Italy}

\author{P. Homm}
\affiliation{Department of Physics and Astronomy, KU Leuven, Celestijnenlaan 200D, 3001 Leuven, Belgium}

\author{M. Menghini}
\affiliation{Department of Physics and Astronomy, KU Leuven, Celestijnenlaan 200D, 3001 Leuven, Belgium}
\affiliation{IMDEA Nanociencia, Cantoblanco, 28049, Madrid, Spain}

\author{P. Franceschini}
\affiliation{Department of Physics, Universit\`a Cattolica del Sacro Cuore, Brescia I-25121, Italy}
\affiliation{ILAMP (Interdisciplinary Laboratories for Advanced Materials Physics), Universit\`a Cattolica del Sacro Cuore, Brescia I-25121, Italy}
\affiliation{Department of Physics and Astronomy, KU Leuven, Celestijnenlaan 200D, 3001 Leuven, Belgium}

\author{F. Maccherozzi}
\affiliation{Diamond Light Source, Chilton, Didcot, Oxfordshire, OX11 0DE, UK}

\author{F. Banfi}
\affiliation{Department of Physics, Universit\`a Cattolica del Sacro Cuore, Brescia I-25121, Italy}
\affiliation{ILAMP (Interdisciplinary Laboratories for Advanced Materials Physics), Universit\`a Cattolica del Sacro Cuore, Brescia I-25121, Italy}
\affiliation{Universit\'e de Lyon, Institut Lumi\`ere Mati\`ere (iLM), Universit\'e Lyon 1 and CNRS
10 rue Ada Byron, 69622 Villeurbanne cedex, France}

\author{G. Ferrini}
\affiliation{Department of Physics, Universit\`a Cattolica del Sacro Cuore, Brescia I-25121, Italy}
\affiliation{ILAMP (Interdisciplinary Laboratories for Advanced Materials Physics), Universit\`a Cattolica del Sacro Cuore, Brescia I-25121, Italy}

\author{F. Cilento}
\affiliation{Elettra-Sincrotrone Trieste S.C.p.A., 34149 Basovizza, Italy}

\author{F. Parmigiani}
\affiliation{Elettra-Sincrotrone Trieste S.C.p.A., 34149 Basovizza, Italy}
\affiliation{Dipartimento di Fisica, Universit\`a degli Studi di Trieste, 34127 Trieste, Italy}
\affiliation{International Faculty, University of Cologne, Albertus-Magnus-Platz, 50923 Cologne, Germany}

\author{S.S. Dhesi}
\affiliation{Diamond Light Source, Chilton, Didcot, Oxfordshire, OX11 0DE, UK}

\author{M. Fabrizio}
\affiliation{Scuola Internazionale Superiore di Studi Avanzati (SISSA), Via Bonomea 265, 34136 Trieste, Italy}

\author{J.-P. Locquet}
\affiliation{Department of Physics and Astronomy, KU Leuven, Celestijnenlaan 200D, 3001 Leuven, Belgium}

\author{C. Giannetti}
\email{claudio.giannetti@unicatt.it}
\affiliation{Department of Physics, Universit\`a Cattolica del Sacro Cuore, Brescia I-25121, Italy}
\affiliation{ILAMP (Interdisciplinary Laboratories for Advanced Materials Physics), Universit\`a Cattolica del Sacro Cuore, Brescia I-25121, Italy}

\begin{abstract}
Unveiling the physics that governs the intertwining between the nanoscale self-organization and the dynamics of insulator-to-metal transitions (\textit{IMT}) is key for controlling on demand the ultrafast switching in strongly correlated materials and nano-devices.
A paradigmatic case is the \textit{IMT} in V$_2$O$_3$, for which the mechanism that leads to the nucleation and growth of metallic nano-droplets out of the supposedly homogeneous Mott insulating phase is still a mystery. 

Here, we combine X-ray photoemission electron microscopy and ultrafast non-equilibrium optical spectroscopy to investigate the early stage dynamics of isolated metallic nano-droplets across the \textit{IMT} in V$_2$O$_3$ thin films. Our experiments show that the low-temperature monoclinic antiferromagnetic insulating phase is characterized by the spontaneous formation of striped polydomains, with different lattice distortions. The insulating domain boundaries accommodate the birth of metallic nano-droplets, whose non-equilibrium expansion can be triggered by the photo-induced change of the 3$d$-orbital occupation. We address the relation between the spontaneous nanotexture of the Mott insulating phase in V$_2$O$_3$ and the timescale of the  metallic seeds growth. We speculate that the photoinduced metallic growth can proceed along a non-thermal pathway in which the monoclinic lattice symmetry of the insulating phase is partially retained. 
\end{abstract}

\maketitle
\section{Introduction}
Insulator-to-metal transitions (\textit{IMT}) in strongly correlated materials are among the most remarkable examples of solid-solid transformations driven by electronic interactions \cite{Imada1998}. The prototypical Mott insulator V$_2$O$_3$ \cite{McWhan1970}, which undergoes a first-order \textit{IMT} from a monoclinic antiferromagnetic insulator to a corundum paramagnetic metal, is widely considered a unique platform to investigate the \textit{IMT} dynamics and benchmark out-of-equilibrium models for correlated materials.
On one hand, the recent development of optical and photoemission spectroscopies with spatial resolution unveiled an inherent inhomogeneity of the \textit{IMT} \cite{Qazilbash2007,Lupi2010,OCallahan2015,Lantz2016,McLeod2017,Vidas2018}, which suggests the existence of a hidden nanotextured template in the insulating phase that drives the transition dynamics. On the other hand, time-resolved techniques have been used to investigate the \textit{IMT} dynamics triggered by impulsive excitation with light pulses intense enough to overcome the energy barrier due to the latent heat and induce the structural phase transition \cite{Singer2018} and the emergence of mesoscopic metallicity \cite{Liu2011,Abreu2015}. However, the early stages of the formation of nanometric metallic seeds within the Mott insulating phase and the subsequent growth that finally leads to the formation of percolative mesoscopic metallic networks still remain a no man's land.  
Unveiling the dynamics of nascent metallic nano-droplets at the relevant space- and time-scales would provide fundamental insights on the forces driving the phase transformation and represent a challenge for models of the electrical breakdown of Mott insulators, phenomenologically described by resistor network models \cite{Stoliar2013}. In this picture, each node is assumed to represent a nanometric patch that transforms from insulating to metallic in the presence of an above-threshold electric field. The resistive transition occurs when a connected filamentary path of metallized nodes forms. Addressing the physical meaning of the nodes and the precise mechanisms by which they turn metallic is key to control the ultrafast \textit{IMT} and the electrical breakdown in nanometric resistive-switching devices capable of operating at THz frequencies \cite{Stoliar2013,Basov2017,Tokura2017}.

\begin{figure*}
\includegraphics[keepaspectratio,clip,width=0.9\textwidth]{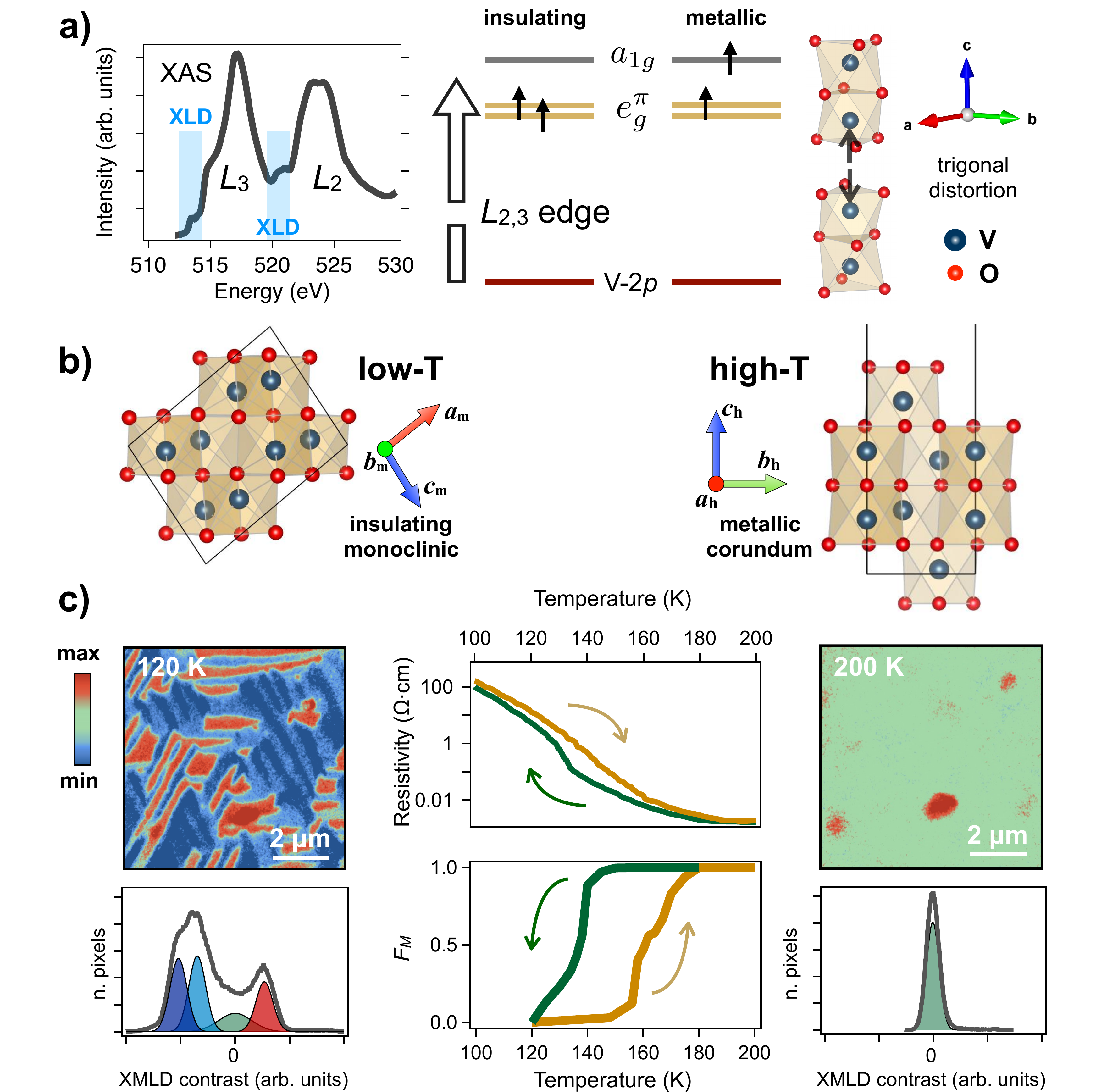}
\caption{\textbf{Nanoscale insulating domains imaged by XLD-PEEM.} \textbf{a)} Left: X-ray Absorption Spectroscopy (XAS) spectrum showing the $L_{2,3}$ vanadium absorption lines. The light blue areas highlight the spectral regions where the maximum XLD signal is obtained. Middle: Sketch of the $L_{2,3}$ transitions from the V-2$p$ levels to the doubly occupied V-$t_{2g}$ levels further split into $a_{1g}$ and $e^{\pi}_g$ by the trigonal distortion of the lattice. Right: trigonal distortion of the V atoms along the $c$-axis of the hexagonal unit cell. \textbf{b)} Left: $a$-$c$ plane projection of the monoclinic lattice structure in the low-$T$ AFI phase. The $a_\mathrm{m}$, $b_\mathrm{m}$ and $c_\mathrm{m}$ axes of the monoclinic cell are indicated by the colored arrows.  Right: $c$-$b$ plane projection of the corundum lattice structure in the high-$T$ PM phase. The $a_\mathrm{h}$, $b_\mathrm{h}$ and $c_\mathrm{h}$ axes of the hexagonal cell are indicated by the colored arrows.  \textbf{c)} Left top: XLD-PEEM image taken at 120 K (heating branch of the hysteresis) evidencing striped monoclinic domains. Left bottom: XLD intensity distribution (thick grey line) of the image. The analysis evidences a broad zero-centered distribution (green), corresponding to the resolution-broadened edges of the striped domains, and three distinct distributions with non-zero signal (red, blue, light blue), reflecting the different structural nanodomains within the AFI  phase. Middle: hysteresis cycle as obtained by resistivity measurements and by the metallic filling analysis of the temperature-dependent XLD-PEEM images (see Appendix D). The yellow (green) arrows indicate the heating (cooling) cycles. Right top: XLD-PEEM image taken at 200 K (heating branch of the hysteresis) showing a homogeneous background, typical of the PM phase. The residual red spots are due to defects used as reference to precisely align the XLD-PEEM images. Right bottom: XLD intensity distribution (thick grey line) of the image. The analysis evidences a zero-centered single-mode distribution.} 
\label{Fig_general}
\end{figure*}

Here, we combine element-specific X-ray nanoimaging with ultrafast broadband optical spectroscopy to investigate the early stages of the formation of nanometric metallic seeds across the \textit{IMT} in V$_2$O$_3$ thin films. We discover an intrinsic nanotexture of the monoclinic antiferromagnetic insulating (AFI) phase, which is characterized by striped and ordered polydomains with different lattice distortions. In the insulator-metal coexisting region, the boundaries of the insulating stripes seed the formation of metallic nuclei, whose proliferation leads to the formation of mesoscopic metallic networks through a percolative \textit{IMT} transition. 
The growth dynamics of the metallic seeds is investigated by probing the transient optical conductivity in the 1.4-2.2 eV energy range, which is extremely sensitive to the local bandstructure rearrangement in the nanometric metallic regions. We demonstrate that weak light excitation, i.e. with fluence smaller than that necessary to overcome the latent heat and photoinduce a mesoscopic \textit{IMT}, triggers the growth of the metallic filling factor in the system. The dynamics saturates after 25-30 ps, which corresponds to the time necessary for the propagation at the sound velocity of metallic droplets across the $\sim$250 nm wide insulating stripes. 

\section{Imaging the nanotexture of the monoclinic insulating phase via X-ray microscopy}

The low-energy physics of vanadium sesquioxide (V$_2$O$_3$) originates from the interactions between the two electrons occupying the V-3$d$ levels and from their coupling to the lattice deformations. In particular, the octahedral crystal-field  splits the V-3$d$ orbitals into lower $t_{2g}$ and upper $e_g$ bands. Since the octahedra display a further trigonal distortion, reinforced by the on-site Coulomb repulsion ($U\simeq$2.5 eV \cite{Qazilbash2008}), the $t_{2g}$ orbitals are split (see Fig. \ref{Fig_general}a) in turn into a lower $e^{\pi}_{g}$ doublet, mainly oriented in the $a$-$b$ plane, and an upper $a_{1g}$ singlet. In the low-$T$ AFI  monoclinic phase, the average occupancy of $e_{g}^{\pi}$ orbitals ($n_{e_{g}^{\pi}}\simeq$0.83) largely overcomes that of the $a_{1g}$ level ($n_{a_{1g}}\simeq$0.17) and the system behaves as a half-filled two-band Mott insulator \cite{Park2000} with an effective charge gap of 2$\Delta_{\mathrm{eff}}\simeq$0.5 eV \cite{Qazilbash2008}. At $T_{IMT} \sim$160 K, the first-order structural change from monoclinic to corundum (see Fig. \ref{Fig_general}b), which implies an increase of crystal symmetry and a reduction of the unit cell volume, accompanies a jump of the $a_{1g}$ occupation with a concomitant transition to the paramagnetic metallic (PM) phase, characterized by  $n_{a_{1g}}\simeq$0.25 \cite{Park2000}.
Across the \textit{IMT}, the electrical resistivity can be tuned over several orders of magnitude by means of small variations in chemical composition \cite{Kuwamoto1980,Homm2015,delValle2017}, pressure \cite{McWhan1969,Jayaraman1970,Limelette2003,Valmianski2017}, temperature \cite{McLeod2017} and, more recently, by external application of electric fields \cite{Guenon2013,Stoliar2013,Mazza2016} or irradiation with short light pulses \cite{Mansart2010,Liu2011,Abreu2015,Abreu2017,Lantz2017,Giorgianni2019}.

\begin{figure}
\includegraphics[keepaspectratio,clip,width=0.6\textwidth]{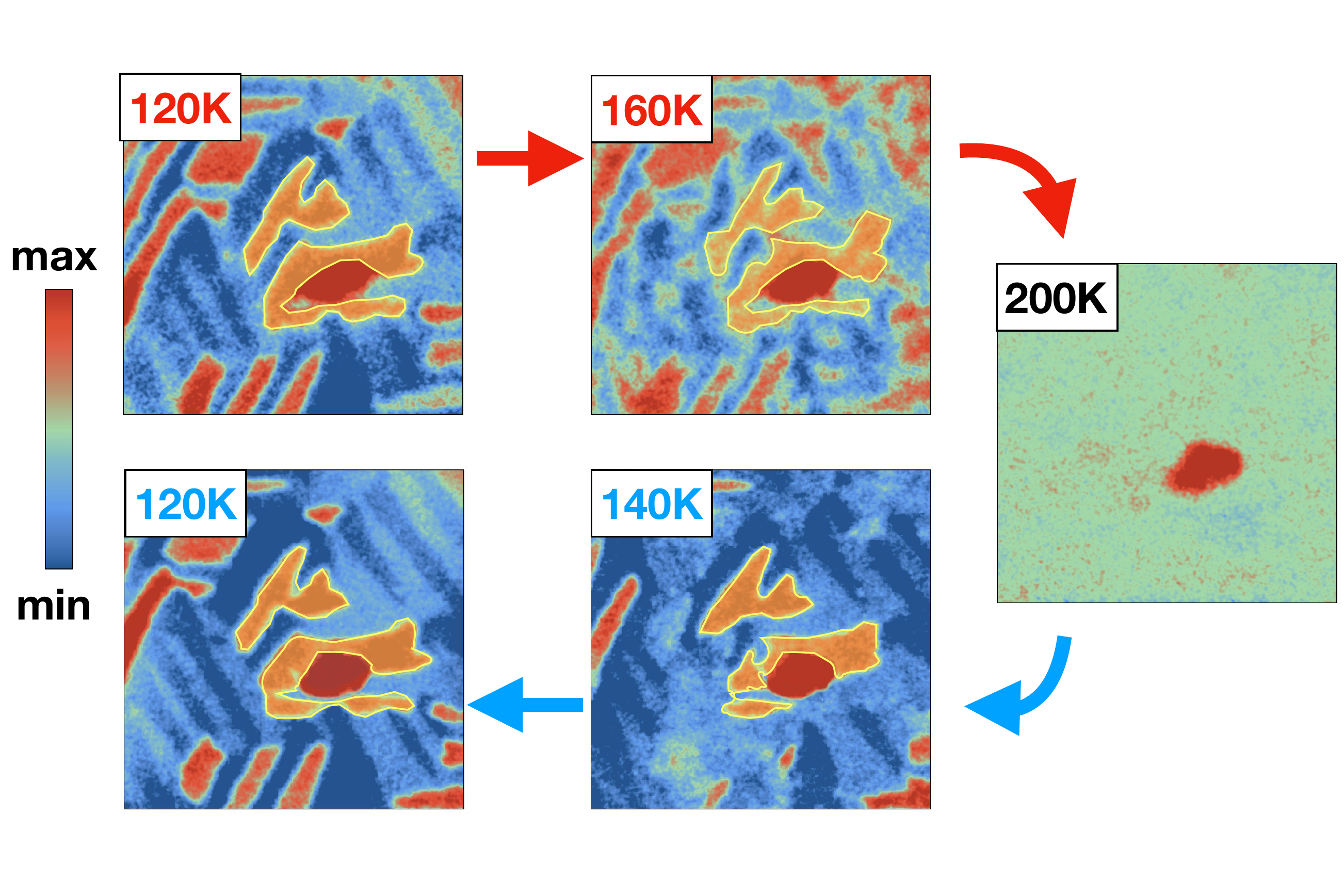}
\caption{\textbf{Domain pinning and hysteresis.} XLD-PEEM images at different temperatures during a heating and cooling cycle. After the complete melting of the AFI phase at $T$=200 K, the insulating striped domains forms in the same spatial position when the system is cooled down to $T$=120 K.} 
\label{Fig_cycle}
\end{figure}

\begin{figure*}
\includegraphics[keepaspectratio,clip,width=1\textwidth]{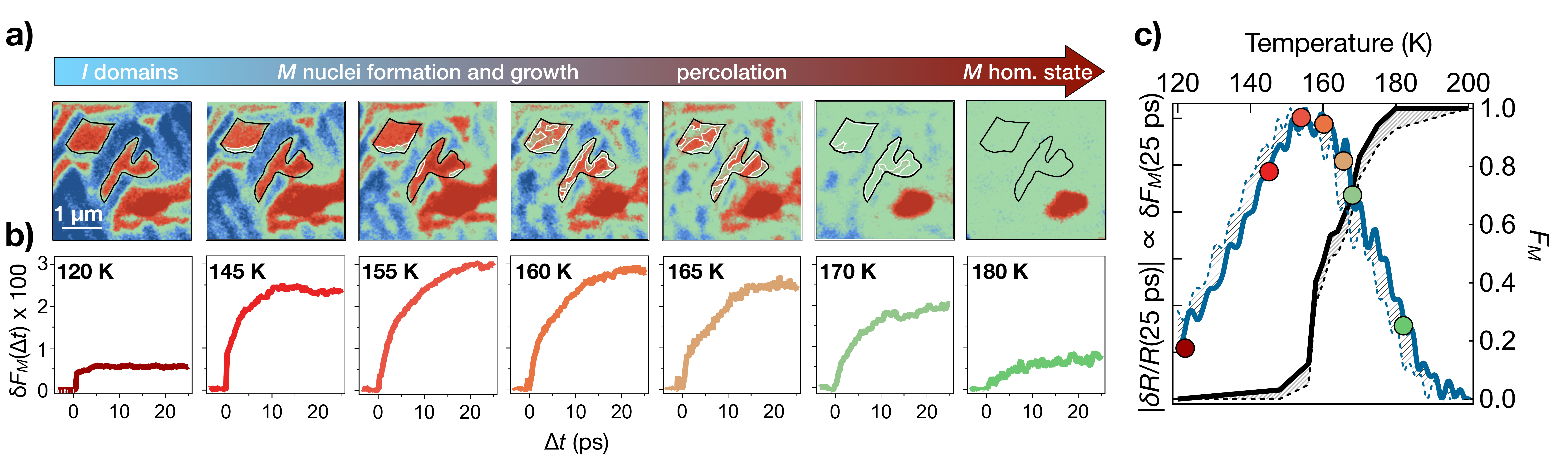}
\caption{\textbf{Growth dynamics of metallic seeds.}  \textbf{a)} Temperature-dependent XLD-PEEM images. The color scale is the same as that reported in Fig. \ref{Fig_general}c. The black contours highlight the monoclinic domain on which the metallic filling analysis has been performed. The white contours indicate the metallic regions within the monoclinic domains as identified via the routine described in the Appendix D. The arrow highlights the main processes leading to the complete insulator-to-metal phase transition. \textbf{b)} Pump-induced filling factor variation, $\delta F_M(\Delta t)$, at different temperatures below (120 K), within (145-165 K) and above ($>$180 K) the $I$-$M$ coexistence region. The measurements have been performed using setup SU2, as described in appendix C, with an incident pump fluence of 0.3 mJ/cm$^2$.  \textbf{c)} Absolute value of the long-time ($\Delta t$=25 ps) reflectivity variation as a function of the temperature (blue line, left axis). The measurements have been performed using setups SU2 and SU3, as described in appendix C. The maximum reflectivity variation, $\delta R/R$(25 ps)$\simeq$2.2$\cdot10^{-3}$ at 2.1 eV photon energy, is measured at $T\simeq$155 K and corresponds to a filling factor variation $\delta F_M\simeq$3\%. The dashed area between the solid and dashed blue lines indicates the uncertainty ($\sim$4 K) in the actual sample temperature during the fast ramping up ($\sim$2 K/minute) of the temperature of the sample holder. For sake of comparison, the black line represents the metallic filling in the heating cycle as extracted from the XLD-PEEM images (see Fig. \ref{Fig_general}c). The small mismatch between the temperatures at which $\delta R/R$ vanishes ($\sim$190 K) and $F_M\simeq$1 ($\sim$180 K) is related to the finite resolution of the XLD-PEEM that is insensitive to insulating domains smaller than $\sim$30 nm. The grey dashed area between the solid and dashed black lines corresponds to the estimated  uncertainty of $F_M$ due to the finite PEEM resolution. See Supplementary Information for a detailed discussion of the role of PEEM resolution in retrieving the $F_M$ value from the pixel-segmentation analysis of the XLD-PEEM images. The colored dots indicate the temperatures of the XLD-PEEM images and time-traces as reported in panel b).} 
\label{Fig_dynamics}
\end{figure*}

Snapshots of the \textit{IMT} in real space have been taken via photoemission electron microscopy (PEEM) on a $\SI{40}{nm}$ V$_2$O$_3$ film epitaxially grown on a (0001)-Al$_2$O$_3$ substrate, therefore with the $c$-axis oriented perpendicular to the surface of the film (see Appendix A). PEEM images have been obtained by exploiting X-ray Linear Dichroism (XLD) \cite{Scholl2000} at the V-$L_{2,3}$-edge ($\sim$513-530 eV) as the contrast mechanism. In the AFI phase, the XLD signal at the specific energies of 513 and 520 eV arises from the absorption difference between X-rays polarized parallel and perpendicular to the $a_{1g}$ orbitals \cite{Park2000}, which have lobes oriented along the $c$-axis. The different contrast between the AFI and PM phases originates from the different occupancy ratio, $n_{a_{1g}}/n_{e^{\pi}_g}$, within the V-$t_{2g}$ orbitals \cite{Park2000,Rodolakis2010}. In Fig. \ref{Fig_general}c (top-left) we report the spatially-resolved XLD contrast at 513 eV in the AFI phase ($T$=120 K). Striped insulating nano-domains, oriented along the hexagonal crystallographic axes, are clearly visible with characteristic dimensions of a few micrometers in length and 200-300 nm in width. The analysis of the intensity distribution (see Appendix D) highlights the presence of three different domains, which correspond to three possible orientations of the monoclinic domains subject to distortions along the three equivalent edges of the hexagonal unit cell \cite{Dernier1970,Singer2018}. 
When the temperature is increased, the XLD contrast of the insulating domains progressively vanishes until the uniform metallic phase, characterized by a single color distribution, emerges (see right-top panel in Fig. \ref{Fig_general}c). The metallic filling $F_M$ is calculated as the ratio between the number of metallic pixels (green XLD contrast), recognized by pixel-based segmentation techniques (see Appendix D), and the total number of pixels within a fixed monoclinic domain. The temperature dependence of $F_M$ is shown in the bottom-middle panel of Fig. \ref{Fig_general}c. The metallic filling displays a large temperature hysteresis ($\delta T\sim$20 K), which corresponds to the resistivity hysteresis measured on the same sample and reported in the top-middle panel of Fig. \ref{Fig_general}c.
We point out that, under temperature cycling, the striped domains always nucleate at the same location and with the same characteristics (see Fig. \ref{Fig_cycle}), thus suggesting that the formation of nanotextured polydomains is not only unavoidable in the transformation from the high temperature corundum structure to the low temperature monoclinic one \cite{Roytburd1995}, but also reproducible.
When focusing on specific monoclinic domains (see black contours in the XLD-PEEM images shown in Fig. \ref{Fig_dynamics}a) we observe the emergence of finite-size metallic nuclei, i.e. metallic regions with a critical width larger than the experimental resolution of 30 nm, at temperatures above 145 K. The metallic phase nucleates along the boundaries between different monoclinic domains, which is a typical characteristic of martensitic transitions \cite{Papon}. Martensitic effects in the course of the $IMT$ in V$_2$O$_3$ were already suggested on the basis of acoustic emission experiments \cite{Chudnovskii1997}. When further increasing the temperature, the growth of the elongated metallic nuclei leads to a percolative transition \cite{McLeod2017} in a narrow temperature range centered at $\sim$160 K, which corresponds to $F_M^{perc}$=0.45. At 180 K, the average dimension of the residual insulating puddles is below the experimental resolution and the metallic phase eventually occupies most of the pristine insulating domains ($F_M\simeq$1). 

\section{Growth dynamics of metallic droplets embedded in the nanotextured insulating phase}
\subsection{Dynamics triggered by photoinduced orbital polarization change}
The intrinsic nano-texture of the insulating phase unveiled by the XLD-PEEM experiment constitutes a natural template, which regulates the metallic nuclei formation and possibly affects their growth and expansion. The picosecond dynamics of the increase of metallic filling can be directly investigated in the time-domain by time-resolved optical spectroscopy, which employs ultrashort light pulses to impulsively weaken the AFI phase and drive the temporal evolution of the metallic seeds. The excitation process can be understood starting from the following thermodynamic considerations.    
At the equilibrum temperature $T$, the distribution function of stable metallic nuclei, i.e. those with volume of the order of the critical volume $v_{\mathrm{cr}}$, is $f(v_{\mathrm{cr}}) \propto$exp($-G_{cr}/kT$), where $G_{cr}$ is the energy barrier necessary to form a stable nucleus. This energy barrier, $G_{cr}$=($g_M-g_I$)$v_{\mathrm{cr}}$+$\alpha s_{\mathrm{cr}}$, is a balance between the volume energy gain to form metallic nuclei ($g_M<g_I$) and the surface energy loss associated with the surface tension ($\alpha$) \cite{Landau,Papon} and  critical surface area ($s_{\mathrm{cr}}$). The effect of the excitation with light pulses shorter ($\sim$50 fs) than the typical dynamics of the phase transformation and with photon energy ($\hbar\omega$=1.55 eV) larger than the charge-gap is thus twofold. On one hand, the energy absorbed is rapidly released to the lattice and leads to a quick increase ($\delta T$) of the local effective temperature within less than one picosecond \cite{Mansart2010,Lantz2017}. On the other hand,
the impulsive photoexcitation modifies the population of the $e_{g}^{\pi}$ and $a_{1g}$ orbitals (see Fig. \ref{Fig_optics}), which constitutes the control parameter of the \textit{IMT} \cite{Sandri2015,Lantz2017,Ronchi2018}. The light thus induces a change in the orbital polarization, $\delta p$=$\delta (n_{e^{\pi}_g}-n_{a_{1g}})$ (see Fig. \ref{Fig_optics}a), that weakens the Mott insulating phase and drives a further decrease of the free energy density difference between the insulating and metallic phases, as given by \cite{Sandri2015,Ronchi2018}:
\begin{equation}
\delta (g_M-g_I)\sim -\frac{U}{2}\delta p
\end{equation}
The light-induced increase of $T$ and the simultaneous change of the orbital population is the microscopic mechanism that, in the $I$-$M$ coexisting region, triggers the generation and growth of isolated metallic nuclei within the striped insulating domains. At high excitation fluence, the distribution of metastable metallic nuclei will eventually overcome the percolative threshold
, thus leading to the formation of connected metallic networks and the emergence of mesoscopic conductivity.

\subsection{Time-resolved reflectivity measurements in the low-fluence regime}
Optical pump-THz probe experiments have been widely used \cite{Liu2011,Abreu2015} to probe the dynamics of mesoscopic metallization. When the absorbed pump energy overcomes the threshold necessary for inducing the \textit{IMT}, the onset of a finite THz conductivity variation has been measured on a timescale of  50-100 ps \cite{Abreu2015}. 
Although THz light is a direct probe of metallicity, the presence of domains smaller than the characteristic transport length  scale, $L$=$\sqrt {D/\omega}$ where $D$ is the diffusion coefficient, strongly affects the low-frequency conductivity \cite{Henning1999}. Therefore, THz properties do not exclusively depend on the metallicity, but also on the connectivity among the different metallic domains, which dramatically changes across the percolative transition \cite{Henning1999,Walther2007}. As a direct consequence, the modeling of the low-frequency optical conductivity across the \textit{IMT} in VO$_2$ and V$_2$O$_3$ requires a modified Drude model, which accounts for carrier localization in nanograins and directional scattering at the crossover between ballistic and diffusive electronic motion \cite{Jepsen2006,Cocker2010,Luo2017}. 
In order to probe the early dynamics of isolated metallic nanograins below the threshold necessary to induce mesoscopic metallization we use optical spectroscopy in the infrared-visible range. As a consequence of the reduced characteristic transport length, visible light constitutes an extremely sensitive probe to bandstructure changes localized in isolated nanometric spatial regions, independently of the formation of percolative patterns. We also note that across the insulator-to-metal Mott transition the spectral weight of the optical transitions involving the Hubbard bands mirrors the low-energy spectral weight of the Drude conductivity. The V$_2$O$_3$ optical properties in the visible range are dominated by the transition between the $e^{\pi}_g$ lower (LHB) and upper Hubbard bands (UHB) \cite{Qazilbash2008,Stewart2012}, which is accounted for by a broad Lorentz oscillator centered at 2 eV (see Supplementary Information). In Fig. \ref{Fig_optics}b and c, we report the equilibrium reflectivity variation measured at selected photon energies during the heating cycle, i.e. $[R(T,\omega)$-$R_M(\omega)]$/$R(T,\omega)$, where $R(T,\omega)$ is the generic temperature-dependent reflectivity and $R_M(\omega)$ is the reflectivity of the metallic phase, taken as the average of $R(T,\omega)$ between 190 K and 250 K. Starting from the reflectivity of the insulating phase at $T$=120 K, i.e. $R$(120 K,$\omega$)=$R_I(\omega)$, a maximum drop of $\sim$8\% across the \textit{IMT} is measured in the 1.8-2.1 eV photon energy range, which corresponds to the $e^{\pi}_g$ LHB$\rightarrow$UHB transition. We can therefore conclude that the spectral weight variation at optical frequencies (1.4-2.2 eV) is a direct measure of the change of the metallic filling fraction and that broadband pump-probe reflectivity measurements can be used to snap the pump-induced growth dynamics of isolated metallic grains even at fluences much smaller than those necessary to overcome the latent heat and induce mesoscopic metallicity.
More quantitatively, the pump-induced variation of the effective dielectric function, $\epsilon_{eff}$, of an inhomogeneous phase can be modeled by the Bruggeman effective medium approximation (BEMA) (see Supplementary Information), which correctly describes the percolative \textit{IMT} in vanadates \cite{Bruggeman1935,Jepsen2006,Lupi2010}. In general, for small filling factor variations ($\delta F_M$), the transient reflectivity change between the excited and equilibrium states can be linearized as follows:
\begin{equation}
\frac{\delta R}{R}(F_M,\omega,\Delta t)=\frac{R_{\mathrm{exc}}(F_M,\omega,\Delta t)-R_{\mathrm{eq}}(F_M,\omega)}{R_{\mathrm{eq}}(F_M,\omega)}\simeq\Phi(F_M,\omega)\delta F_M(\Delta t)
\label{eq_BEMA}
\end{equation}
where $\Delta t$ is the delay between the pump and probe pulses and $\Phi(F_M,\omega)$ is a function of both the initial filling factor and the probe frequency. In the Supplementary Information, we demonstrate that $\Phi(F_M,\omega)\simeq[R_M(\omega)$-$R_I(\omega)]$/$R_I(\omega)]$ in the 1.4-2.2 eV region and for any initial $F_M$ value.
As a consequence, we obtain that $\delta F_M(\Delta t)\simeq\delta R/R(\omega,\Delta t)\cdot R_{I}(\omega)/[R_{M}(\omega)$-$R_{I}(\omega)]$, where $[R_{M}(\omega)$-$R_{I}(\omega)]$/$R_{I}(\omega)$ is the relative reflectivity difference between the metallic and insulating phases, as obtained from the equilibrium reflectivity.  
In Fig. \ref{Fig_optics}c we report the $\delta R/R(\omega,\Delta t)$ signal measured at $T$=145 K ($F_M$=0.01) as a function of the probe photon energy and at fixed delay time ($\Delta t$=25 ps). The frequency dependence of the transient reflectivity variation perfectly reproduces the relative difference between the reflectivities measured at the equilibrium temperatures $T$=145 K and $T$=200 K (see Fig. \ref{Fig_optics}b), thus experimentally proving that $\frac{\delta R}{R}(\omega,\Delta t)\propto [R_M(\omega)]$-$R_I(\omega)]$/$R_I(\omega)]$. The range of validity of Eq. \ref{eq_BEMA} extends to $\delta F_M$ values as large as 30\%, as shown in the Supplementary Information. On the other hand, considering that the photoinduced $\delta R/R(\omega,\Delta t)$ signal corresponds to a few percent of the static $[R_M(\omega)]$-$R_I(\omega)]/R_{I}(\omega)$ difference and is measured with a signal-to-noise ratio of 10$^{-6}$, pump-induced changes of $F_M(\Delta t)$ as small as 10$^{-4}$ are well within the experimental capability.

\begin{figure}
\includegraphics[keepaspectratio,clip,width=1\textwidth]{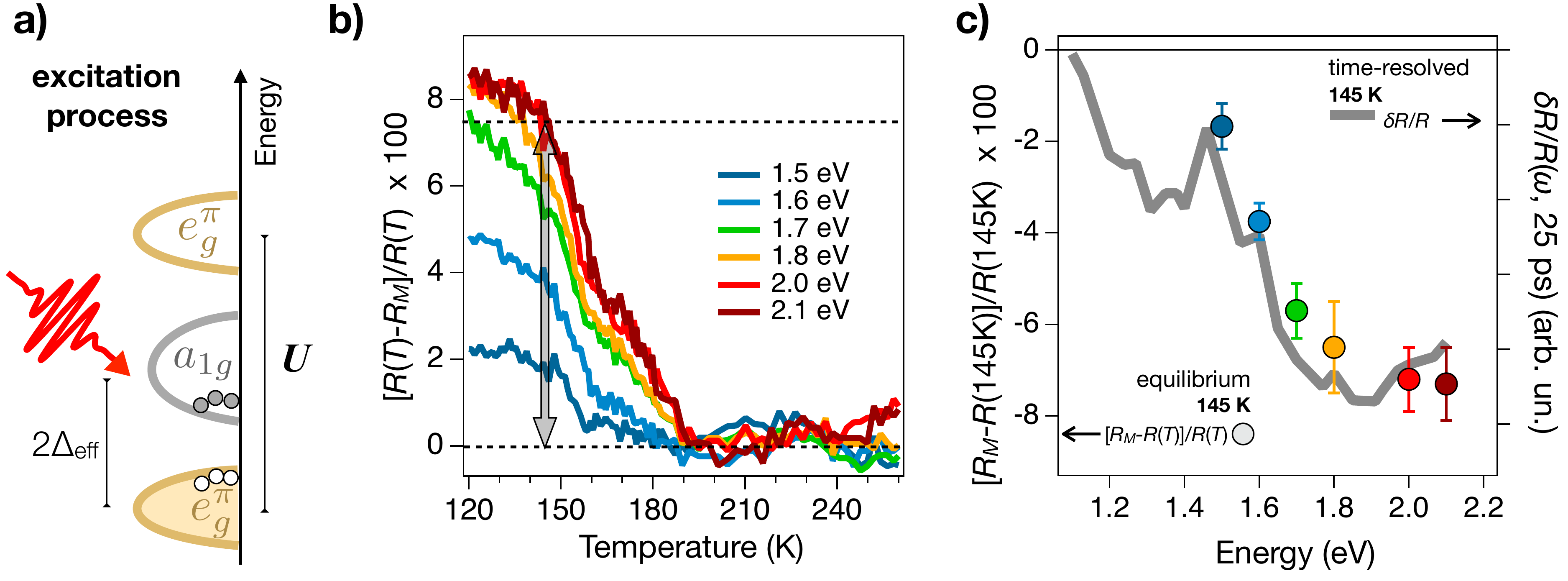}
\caption{\textbf{Energy-resolved differential reflectivity variation.} \textbf{a)} Cartoon of the photoexcitation across the Mott gap. The change in the occupation of the $a_{1g}$ and $e^{\pi}_g$ orbitals gives rise to the orbital polarization change $\delta p$=$\delta$($n_{e^{\pi}_g}$-$n_{a_{1g}}$), which triggers the growth of metallic seeds. \textbf{b)} Equilibrium infrared/visible reflectivity change, $[R$($T,\omega$)-$R_M(\omega)]$/$R$($T,\omega$), as a function of the sample temperature during a heating cycle. $R_M(\omega)$ is taken as the average of $R$($T,\omega$) between 190 and 250 K. The arrow and the dashed lines graphically shows the procedure to obtain the value $[R_M(\omega)$-$R$(145 K, $\omega)]$/$R$(145 K, $\omega$), reported in panel c), for 2 eV (600 nm) photon energy. The measurements have been performed using setup SU3, as described in appendix C. \textbf{c)} Frequency dependence of the equilibrium reflectivity change, $[R_M(\omega)$-$R$(145 K, $\omega)]$/$R$(145 K, $\omega$) (colored dots, left axis), and of the time-resolved $\delta R/R(\omega$) signal (gray line, right axis) measured at $\Delta t$=25 ps and $T$=145 K. The measurements have been performed using setup SU1, as described in appendix C. As reference, we remind that the amplitude of the reflectivity variation measured at $T$=155 K and $\hbar \omega \simeq$2.1 eV is $\delta R/R$(25 ps)$\simeq$2.2$\cdot10^{-3}$.} 
\label{Fig_optics}
\end{figure}
 
To address the connection between the spatial evolution and the temporal dynamics of the metallic seeds,  we compare the outcomes of time-resolved optical spectroscopy and XLD-PEEM measurements performed on the same sample and in the same points of the hysteresis cycle.
In Fig. \ref{Fig_dynamics}a and \ref{Fig_dynamics}b we show the one-to-one correspondence between the real space XLD-PEEM images of the \textit{IMT} and the time-resolved dynamics of $\delta F_M(\Delta t)$ during the heating cycle. In the time-domain, the existence of stable finite-size metallic regions (white contours in XLD-PEEM images of Fig. \ref{Fig_dynamics}a) for $T>$145 K corresponds to a slower build-up dynamics, i.e. to an increase of the time necessary to achieve the saturation of $\delta F_M(\Delta t)$, with respect to the dynamics measured at lower temperatures (120 K). In the $I$-$M$ coexisting region we also observe an increase of the long-time signal ($\Delta t\sim$25 ps), as shown in Fig. \ref{Fig_dynamics}b. In Fig. \ref{Fig_dynamics}c we report the temperature dependence of $\delta R/R$(25 ps), which exhibits a clear maximum corresponding to the hysteresis region observed by XLD-PEEM. All the experiments have been performed with an incident pump fluence of 0.3 mJ/cm$^2$, which corresponds to an absorbed energy $E_{abs}$=13 J/cm$^3$ (see Appendix C). We stress that the absorbed energy is significantly smaller than the threshold necessary to overcome the latent heat associated to the mesoscopic \textit{IMT}, which ranges from $E_{th}\sim$110 J/cm$^3$ at 100-140 K to $E_{th}\sim$30 J/cm$^3$ at 160 K \cite{Liu2011,Abreu2015,Singer2018}. For this small value of absorbed energy, the maximum pump-induced temperature increase is limited to $\delta T\lesssim$4 K (see Appendix C). At the same time, the light excitation induces an orbital polarization $\delta p\sim 3\cdot$10$^{-3}$ (see Appendix C), corresponding to a free energy density variation $\delta(g_M-g_I)\simeq$-4 meV, which largely overcomes the thermal effect $k_{\mathrm{B}}\delta T$. We thus conclude that the observed dynamics of $\delta F_M(\Delta t)$ is triggered by the impulsive change of orbital polarization and is related to the early-stage growth of isolated metallic domains.  

\subsection{Avrami model and ballistic growth of metallic droplets}
The physical picture emerging from the combination of nano-imaging and picosecond dynamics of the \textit{IMT} is the following: in the temperature range in which the average volume of the metallic domains is smaller than $v_{\mathrm{cr}}$, the system is dominated by the formation of fluctuating small metallic nuclei, which are not captured by the static XLD-PEEM images. In this regime, the light excitation drives a rapid ($<$5 ps) increase of the number of sub-critical metallic nuclei, corresponding to $\delta F_M\sim$0.6\% (see Fig. \ref{Fig_dynamics}b). In the $I$-$M$ coexisting region ($T\gtrsim$145 K), the photo-induced orbital polarization $\delta p$ triggers the growth of metallic seeds at the domain boundaries, which overcome $v_{\mathrm{cr}}$ and irreversibly grow until they encounter the opposite edge of the monoclinic stripes. This process leads to the metallization of a small fraction of the insulating domains, such that the increase of the total metallic filling factor is of the order of $\delta F_M\sim$3\% (see Fig. \ref{Fig_dynamics}b). 

\begin{figure*}
\includegraphics[keepaspectratio,clip,width=1\textwidth]{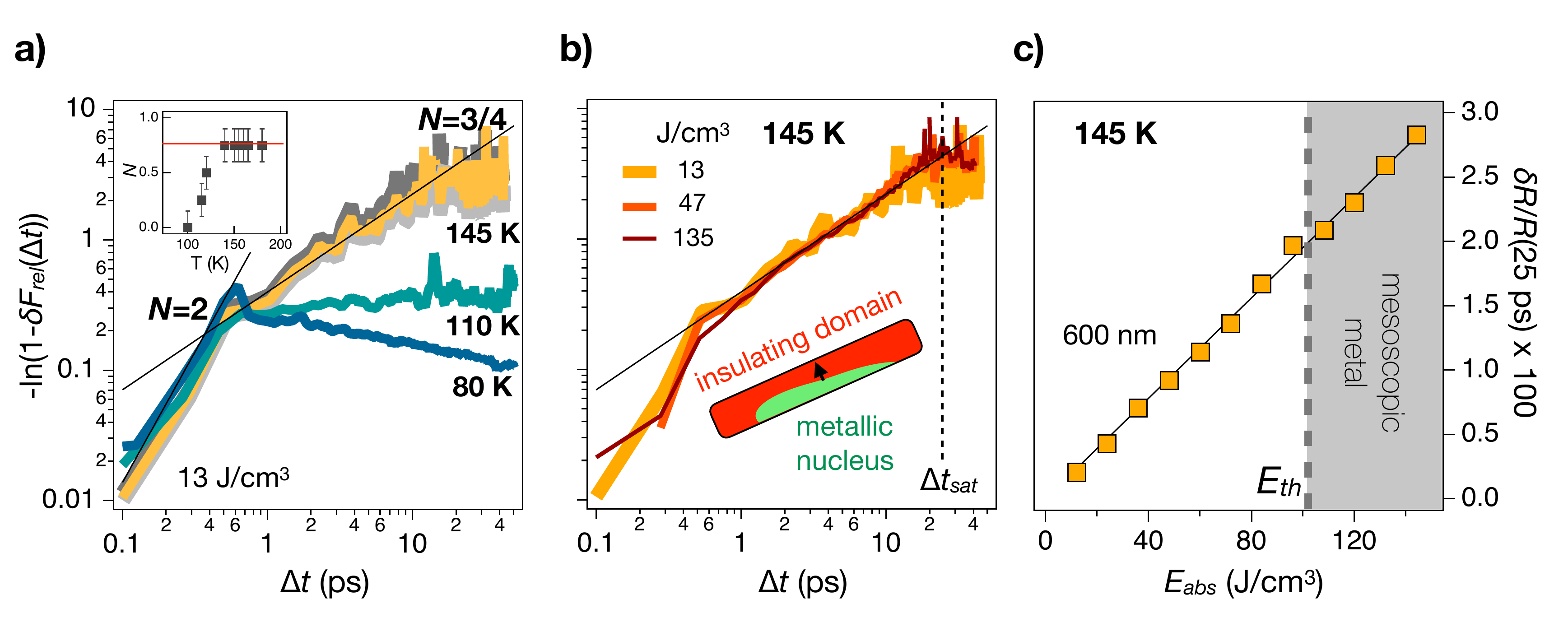}
\caption{\textbf{Avrami model for the metallic growth.} \textbf{a)} Avrami plot of the dynamics of $\delta F_{rel} (\Delta t)$=[$\delta R/R(\Delta t)$]/[$\delta R/R$($\infty$)] at different temperatures. The value of $\delta R/R$($\infty$) is given by the asymptotic value of the exponential fitting to the data in the 0-20 ps time window. The gray traces represent the curves obtained assuming $\delta R/R$($\infty$)$\pm$10\%. The slope does not significantly depend on the value of $\delta R/R$($\infty$). The black lines indicate the expected curves for $N$=2 and $N$=3/4. Inset: Avrami coefficient $N$ of the long-lived dynamics ($\Delta t>$1 ps) as a function of the temperature. The red line indicates the asymptotic value $N$=3/4. \textbf{b)} Avrami plot of $\delta F_{rel}(\Delta t)$ at 145 K for different values of $E_{abs}$. The solid black line indicates the expected curve for $N$=3/4. The dashed line indicates the time at which the growth dynamics saturates. \textbf{c)} The asymptotic value of the relative reflectivity variation, i.e. $\delta R/R$(25 ps), measured at 600 nm wavelength and $T$=145 K, is reported as a function of the absorbed energy density. The black thin line is the linear fit to the data. The dashed line indicates the energy threshold $E_{th}$ above which THz mesoscopic conductivity has been observed \cite{Abreu2015}. The measurements reported in panels a, b and c have been performed using setup SU2, as described in appendix C} 
\label{Fig_avrami}
\end{figure*} 

The nucleation and growth of a new phase at the cost of the initial one is usually described by the Avrami model \cite{Avrami1939,Avrami1940,Avrami1940b}, which predicts that the relative filling factor variation from the equilibrium value $F_M(0)$ to the steady-state $F_M(0)$+$\delta F_M(\infty)$ takes the form: $\delta F_{rel}(\Delta t)$=$\delta F_M(\Delta t)$/$\delta F_M(\infty)$=1-exp(-$K \Delta t^N$), where  $K$ is a constant related to the  growth rate and $N$ is the Avrami coefficient, characteristic of the phase nucleation and growth mode. In Fig. \ref{Fig_avrami} we plot the quantity ln(-ln(1-$\delta F_{rel}(\Delta t)$)), as a function of the logarithmic time ln($\Delta t$). The relative metallic filling is taken as the ratio between the reflectivity variation and its asymptotic value, i.e. $\delta F_{rel} (\Delta t)$=[$\delta R/R(\Delta t)$]/[$\delta R/R$($\infty$)]. At very low temperature ($T$=80 K, Fig. \ref{Fig_avrami}a), we observe a rapid (0-500 fs) growth with $N$=2, followed by a slow decay. While this fast dynamics is compatible with the sudden growth of two-dimensional domains with a rapidly exhausting nucleation rate \cite{Papon}, the slow relaxation to the initial equilibrium condition, i.e. $F_M$=0, indicates that the metallic nuclei never reach the stability threshold. This dynamics is likely associated to the nucleation  of small droplets in the middle of the insulating stripe-like monoclinic domains which are then rapidly reabsorbed. When the temperature is increased, the dynamics of $\delta F_M(\Delta t)$ starts exhibiting the fingerprint of a long-lived growth of the metallic phase, under the form of a linear increase of the signal, in agreement with the prediction of the Avrami law: ln(-ln(1-$\delta F_{rel}(\Delta t)$))=ln($K$)+$N$ln($\Delta t$). The Avrami coefficient progressively increases (see inset in Fig. \ref{Fig_avrami}a) until it reaches a plateau in the $I$-$M$ coexisting region ($T>$145 K), where the data show a linear behavior over more than one decade  (1-20 ps). In this regime the Avrami coefficient ($N$=3/4) is close to unity, as expected for a martensitic transition \cite{Cahn1956}, in which the seeds of the new phase nucleate at the domain boundaries in the first instants and then grow with constant velocity along one direction. 
More specifically, the one-dimensional growth of boundary pinned domains is characterized by $N$=1 for pure interface-controlled growth and by $N$=1/2 for pure diffusion-controlled growth \cite{Papon}. The experimental finding of $N$=3/4 suggests that the interface-controlled dynamics retains some diffusive features, possibly related to the melting of spatially separated metallic droplets.
We note that the simultaneous presence of the $N$=2 and $N$=3/4 dynamics in the time traces measured in the $I$-$M$ coexisting region does not represent a transition between different nucleation and growth regimes, but the simple coexistence of two independent phenomena, which are, respectively, the nucleation of small and unstable metallic droplets in the middle of the insulating monoclinic domains and the one-dimensional irreversible growth of metallic seeds nucleated at the domain boundaries.

The growth of $\delta F_M$ saturates after $\Delta t_{sat}\sim$25-30 ps (see black dashed line in Fig. \ref{Fig_avrami}b) from the impulsive excitation. Considering the sound velocity of V$_2$O$_3$ at 145 K, $v_s\sim$8 km/s \cite{Seikh2006,Abreu2017}, $\Delta t_{sat}$ corresponds to a typical length of 200-240 nm, which is very close to the average width of the striped monoclinic domains observed by XLD-PEEM in the AFI phase. This result suggests that the metallic growth is a ballistic collective process \cite{Abreu2015,Singer2018} in which the metallic domain expansion propagates at the sound velocity, until it encounters the edges of the striped monoclinic domains. Interesting insight on the nature of the non-equilibrium growth of metallic seeds is given by the fluence dependence of  the dynamics. In the $I$-$M$ coexisting region, the $N$=3/4 coefficient is an intrinsic feature, which does not depend on the pump excitation fluence, as shown in Fig. \ref{Fig_avrami}b for excitations densities as high as 135 J/cm$^3$. In Fig. \ref{Fig_avrami}c we also report the asymptotic value of the relative reflectivity variation $\delta R/R$(25 ps) measured at 600 nm wavelength, as a function of the absorbed energy. Surprisingly, $\delta R/R$(25 ps) scales linearly throughout the spanned $E_{abs}$ range and does not exhibit any signature of discontinuity across $E_{th}\sim$110 J/cm$^3$, which is reported as the energy necessary to overcome the latent heat and to observe the onset of metallic THz conductivity \cite{Abreu2015} within 50-100 ps and the mesocopic change of the lattice structure from low-temperature monoclinic to the high-temperature corundum \cite{Singer2018} on similar timescales. 

\begin{figure*}
\includegraphics[keepaspectratio,clip,width=0.8\textwidth]{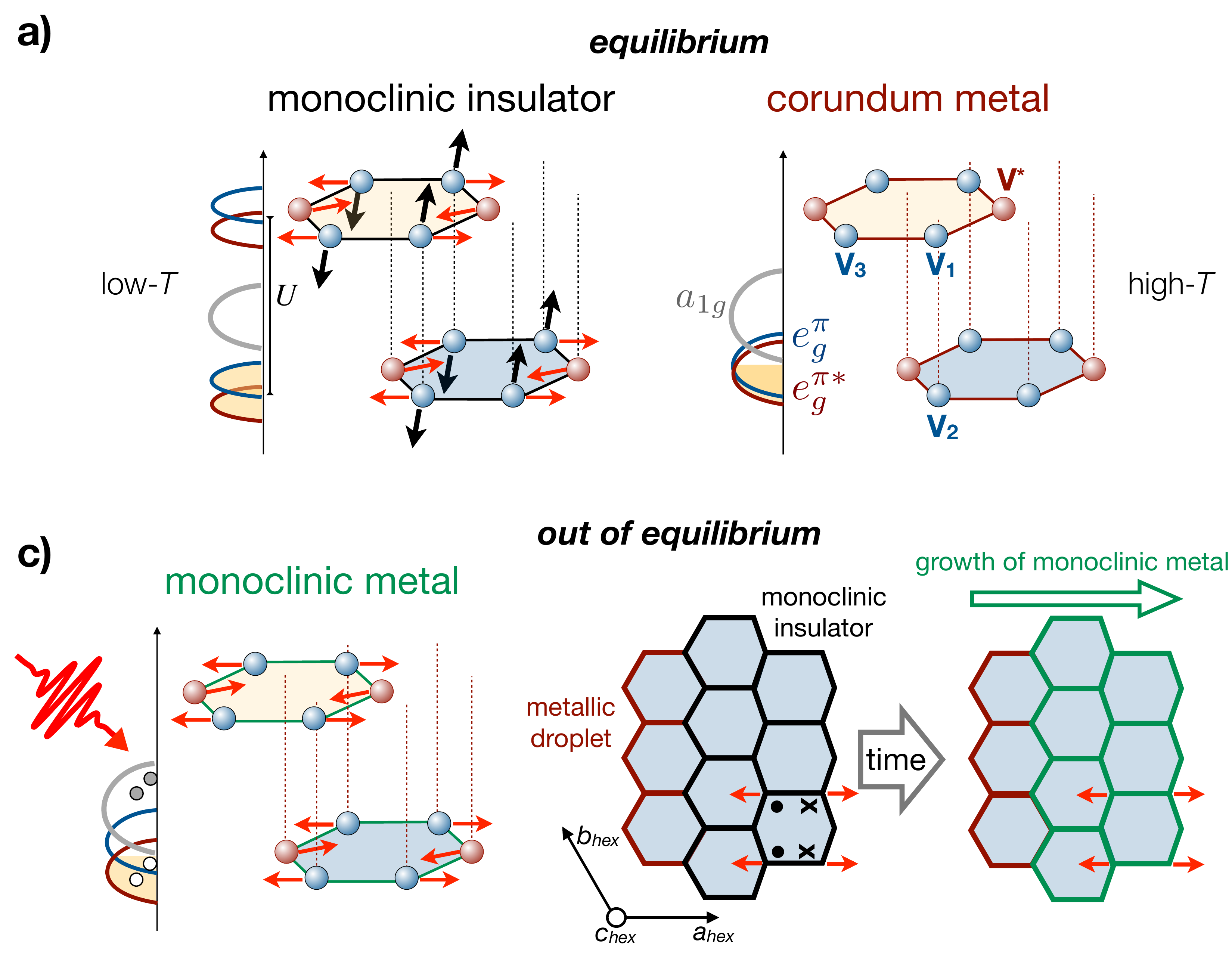}
\caption{\textbf{Cartoon of the non-equilibrium metallic growth dynamics.} \textbf{a)} Sketch of the stacked honeycomb planes in the low-temperature monoclinic and high-temperature corundum phases. In the corundum metal the hexagonal symmetry preserves the equivalence of the V (blue) and V$^*$ (red) sites and, therefore, of the two degenerate $e^{\pi}_g$ and $e^{\pi*}_g$ orbitals. In the monoclinic insulating phase, the distortion of the hexagons (red arrows) lifts the $e^{\pi}_g$ degeneracy, whereas the out-of-plane motion of the vanadium atoms (black arrows), associated to the tilting of the hexagons, controls the $a_{1g}-e^{\pi}_g$ energy distance. \textbf{c)} Sketch of the possible transient metal-like monoclinic state. The photoinduced change of $\delta (n_{e^{\pi}_g}-n_{a_{1g}})$ triggers the restoring of the V$_1$-V$_2$ distance (tilting back of the hexagons), whereas it does not affect the in-plane hexagonal distortion. Photoexcitation thus triggers the growth of already formed metallic droplets via the propagation of a transient non-equilibrium lattice rearrangement (monoclinic metal). The black, red, green borders of the hexagons indicate the monoclinic insulator, the corundum metal and the monoclinic metal, respectively. The red arrows indicate the in-plane distortion of the hexagons, whereas the dots (crosses) indicate the out-of-plane inward (outward) motion of the vanadium atoms associated to the hexagons tilting.} 
\label{Fig_structure}
\end{figure*} 

\section{Discussion}
In order to rationalize the present results, it is instructive to give a closer look to the interplay between the lattice and the electronic bandstructure, as it emerges from calculations accounting for the onsite Coulomb repulsion $U$ \cite{Poteryaev2007,Grieger2015}. If we consider the vanadium atoms, the high-temperature corundum lattice consists in shifted honeycomb planes stacked along the $c$-axis (see Fig. \ref{Fig_structure}a). The hexagonal symmetry of the honeycomb planes preserves the degeneracy of the $e^{\pi}_g$ and ${e^{\pi}_g}^*$ orbitals, which are therefore equivalently occupied. The low-temperature monoclinic antiferromagnetic phase is the result of the breaking of the hexagonal symmetry (see Fig. \ref{Fig_structure}a), associated to the elongation of two edges of the hexagons (V$_1$-V$_3$), which leads to an overall increase of the unit cell volume by 1.4\% \cite{Dernier1970,McWhan1970}. The breaking of the hexagonal symmetry is accompanied by a tilt of the hexagons with respect to the honeycomb planes, which increases the distance of the V$_1$-V$_2$ dimers along the $c$-axis (see Fig. \ref{Fig_structure}a), without further affecting the unit cell volume. While the hexagonal symmetry breaking lifts the degeneracy of the $e^{\pi}_g$ and ${e^{\pi}_g}^*$ orbitals and modifies their population difference, the relative motion of the V$_1$-V$_2$ dimers affects the energy distance between the $e_{g}^{\pi}$ and $a_{1g}$ levels, which controls the metallicity of the system \cite{Poteryaev2007,Grieger2015}. Indeed, metastable metallicity in a lattice structure which retains the monoclinic distortion of vanadium hexagons has been predicted theoretically \cite{Grieger2015} and observed in V$_2$O$_3$ under high pressure \cite{Ding2014} and in VO$_2$ photoexcited \cite{Morrison2014} using a protocol similar to the present experiment.

Our results are compatible with the following intriguing scenario \cite{Grieger2015}. The pump excitation induces a non-thermal increase of the $a_{1g}$ occupation accompanied by the simultaneous depletion of an equal amount of electrons occupying  the two $e^{\pi}_g$ orbitals, i.e. $\delta n_{{e^{\pi}_g}}\simeq\delta n_{{e^{\pi}_g}^*}\simeq \delta n_{a_{1g}}/2$. The photo-excitation thus induces a sudden change of the orbital polarization $\delta p$, which can lead to the transient collapse of the insulating bandstructure \cite{Sandri2015}, but does not significantly alter the population difference between the two $e^{\pi}_g$ orbitals and, as a consequence, does not affect the potential that stabilizes the monoclinic distortion of the vanadium hexagons \cite{Grieger2015}. Within this picture, the pump excitation can trigger the rapid (0-30 ps) proliferation and growth of metallic regions that exhibit transient metal-like electronic properties and a non-thermal lattice structure, obtained by restoring the V$_1$-V$_2$ dimers distance (tilting back of the hexagons) at constant volume while retaining the monoclinic distortion of the vanadium hexagons (see Fig. \ref{Fig_structure}c). Once the metastable metal-like nuclei have filled the striped monoclinic domains, the excess energy stored in the system drives the slower (50-100 ps) first-order transition into the corundum metal \cite{Singer2018}, which involves the restoring of the hexagonal symmetry and the volume decrease of unit cell. This slower process triggers the melting of the isolated metallic grains into a percolative network with mesoscopic metallic properties in the THz \cite{Abreu2015}. The final stage of the metallization thus consists in the two-dimensional growth of mesoscopic metallic puddles, described by the $N$=2 Avrami coefficient, as measured by optical pump-THz probe spectroscopy \cite{Abreu2015}. This multi-step picture of metallic growth is supported by recent time-resolved X-ray diffraction experiments \cite{Singer2018}, that unveiled the fast onset ($\sim$2.5 ps) of a precursor non-thermal structural change of the honeycomb planes triggered by light excitation. This structural dynamics, which is continuous in the excitation energy, preserves the unit cell volume and propagates at the sound velocity. When the excitation energy overcomes the latent heat, the precursory non-thermal lattice dynamics is followed by the mesoscopic transformation into the equilibrium metallic corundum structure.

 
\section{Conclusions}
In conclusion, we shed new light on the intrinsic nanotexture of the monoclinic AFI phase of  V$_2$O$_3$ and its relation with the multi-step nature of the \textit{IMT} dynamics. Exploiting the sensitivity of resonant X-ray PEEM to the orientation of the V-3$d$ orbitals, we unveiled a spontaneous self-organization of the Mott insulating phase, characterized by striped monoclinic domains with different orientations. The insulating stripes are stable and reproducible, as they form in the same position under temperature cycling. Our results show that this nanotexture strongly affects the nucleation and growth dynamics of metallic grains. The insulating domain boundaries host the birth of metallic seeds, whose expansion can be triggered by the impulsive photoinduced change of the orbital population. The timescale of the growth process (25-30 ps) corresponds to the propagation at the sound velocity across the transverse dimension of the insulating stripes. Our data point towards a non-thermal scenario, in which the expanding metallic grains transiently retain the in-plane distortion of the vanadium hexagons typical of the AFI phase.   
The use of XLD-PEEM for the study of nanotextured phases can be readily extended to many different $d$- or $f$-shell correlated materials. The dynamical processes unveiled by our work also impact the modeling of the electrical breakdown in Mott insulators \cite{Stoliar2013}. Our results suggest that each node in the resistor network corresponds to an insulating nanodomain with a well defined orientation of the monoclinic distortion and they show how the node metallizes by light excitation. 
The possibility to suitably engineer the lateral size of the nano-textured insulating domains in thin films is key to control the electronic and magnetic switching of Mott insulators at THz frequencies, with impact on the development of novel Mottronics devices \cite{Tokura2017}.

\section*{Acknowledgments}
C.G. acknowledge financial support from MIUR through the PRIN 2015 Programme (Prot. 2015C5SEJJ001). F.B., G.F. and C.G. acknowledge support from Universit\`a Cattolica del Sacro Cuore through D.1, D.2.2 and D.3.1 grants. The research activities of M.F. have received funding from the European Union, under the project ERC-692670 (FIRSTORM). F.B. acknowledges financial support from the MIUR-Futuro in ricerca 2013 Grant in the frame of the ULTRANANO Project (project number: RBFR13NEA4). P.H., M.M. and J.-P.L. acknowledge support from EU-H2020 Project 688579 PHRESCO. P.H. acknowledges support from Becas Chile-CONICYT. F.P. acknowledge the financial support from UniTS through the FRA 2016 Research Programme. We acknowledge Diamond Light Source for the provision of beamtime under proposal number SI16128.

\subsection*{APPENDIX A: SAMPLES}
The 40 nm epitaxial V$_2$O$_3$ film is deposited by oxygen-assisted molecular beam epitaxy (MBE) in a vacuum chamber with a base pressure of $10^{-9}\ \SI{}{Torr}$. A (0001)-Al$_{2}$O$_{3}$ substrate is used without prior cleaning and is slowly heated to the growth temperature of $\SI{700}{°C}$, as measured with a thermo-couple. V is evaporated from an electron gun with a deposition rate of $\SI{0.1}{\angstrom/s}$ and an oxygen partial pressure of $6.2 \cdot 10^{-6}\ \SI{}{Torr}$ is used during the growth \cite{Dillemans2014}. Under these conditions, a single crystalline film with the $c$-axis oriented perpendicular to the surface is obtained (see Supplementary Material for X-ray diffraction data). 
Temperature dependent resistivity measurements are assessed in the Van der Pauw (VDP) configuration with Au/Cr contacts and using an Oxford Optistat CF2-V cryostat with a sweep rate of $\SI{1.5}{\kelvin}$ per minute.
\subsection*{APPENDIX B: PHOTOEMISSION ELECTRON MICROSCOPY (PEEM)}
PEEM and XAS spectra have been measured at the beamline I06 of Diamond Light Source. The spot size of the X-ray beam is 10 x 10 $\mu$m$^2$. The PEEM images presented in this paper have been obtained by performing XLD experiments at a photon energy of $\sim \SI{513}{eV}$, in proximity of the vanadium $L_{2,3}$ edge. Each PEEM image is the average of the difference between 50 PEEM images taken with light-polarization  parallel (i.e. in the $a$-$b$ plane) and perpendicular (i.e. along the V$_2$O$_3$ $c$-axis) to the film surface, normalized by their sum.  

\subsection*{APPENDIX C: TIME-RESOLVED PUMP-PROBE SPECTROSCOPY}
\label{appendix_time_resolved}
Time-resolved reflectivity measurements  have been performed using three different experimental setups: SU1, a pump supercontinuum-probe \cite{Giannetti2009,Cilento2010}, based on white light (from 0.8 to 3.1 eV, generated by focusing $\sim$ 1 $\mu J$ energy/pulse in a 3 mm thick Sapphire window); SU2, a pump tunable-probe setup, based on an Optical Parametric Amplifier (OPA). In both these cases, a Regenerative Amplifier (Coherent RegA 9000), generating a train of ultrafast pulses (temporal width $\sim$ 50 fs) at $\SI{250}{kHz}$ with a wavelenght $\SI{}{\lambda}=\SI{800}{nm}$ and $\sim$ 6 $\mu J$ energy/pulse, has been used as seed. SU3, a pump-probe setup, based on a Ti:Sapphire cavity-dumped oscillator (Coherent Mira 900), to explore different repetition rates, from 250 kHz to 1 MHz. The results reported do not depend on the repetition rate, thus demonstrating that average heating effects are negligible. Measurements with setups SU1 and SU2 were performed at the T-ReX facility at FERMI, Elettra (Trieste). 
In the setup SU1, the reflected probe beam is spectrally dispersed through an equilateral SF11 prism and imaged on a  Hamamatsu InGaAs linear photodiode array (PDA), capturing the 0.8-2.5 eV spectral region. In configurations SU2 and SU3 the reflected probe is collected by a commercial Si or InGaAs photodiode. For single-color measurements, lock-in acquisition is used, together with fast ($\sim\SI{60} {kHz}$) modulation of the pump beam. This ensures a better signal-to-noise ratio with respect to spectroscopic measurements.
The sample is mounted on the cold finger of a cryostat. The temperature of the sample is stabilized within $\pm\SI{0.5}{\kelvin}$.

In all experiments, the pump photon energy is 1.55 eV with a fluence ranging from $300\, \mu\SI{}{J\cdot cm^{-2}}$ up to  $\SI{3.5}{mJ\cdot cm^{-2}}$. The spot sizes of the pump and probe beams on the sample are $\sim \SI{160}{}\ \mu$m and $\sim \SI{100}{}\ \mu$m in diameter, respectively.
The data reported in Figs. \ref{Fig_dynamics} and \ref{Fig_optics} have been taken with a pump incident fluence of $300\, \mu\SI{}{J\cdot cm^{-2}}$. Considering the lattice specific heat at 160 K, $C_p\sim$3 J cm$^{-3}$K$^{-1}$ \cite{Keer1976}, and estimating the energy absorbed within the 40 nm film via a multi-reflection model (absorption $A$=0.18), we obtain an absorbed energy density $E_{abs}$=13 J/cm$^3$ that corresponds to a temperature increase $\delta T$=$E_{abs}$/$C_p\simeq$4 K. The same absorbed energy density corresponds to an absorbed photon density $n_{ph}\sim$5.4$\cdot$10$^{19}$ cm$^{-3}$, which can be converted into 1.6$\cdot$10$^{-3}$ photons/V atom and an induced orbital polarization $\delta p$=3.2$\cdot$10$^{-3}$. 

\subsection*{APPENDIX D: PEEM IMAGE ANALYSIS}
The analysis of the XLD-PEEM images have been performed by means of the image processing package Fiji \cite{Schindelin2012,Rueden2017}. Whitin this package, the Weka Trainable Segmentation \cite{Arganda2017} machine learning algorithms have been applied to the images to produce pixel-based segmentations, and consequently, to identify different XLD contrast regions. Once the metallic areas inside a single domain have been highlighted, we calculate the metallic filling factor by counting the number of recognized pixel with respect to the total pixels number in a fixed domain. A weighted average is evaluated with respect to the size of several domains to give more accurate results.

\bibliographystyle{apsrev4-1}

\pagebreak
\setcounter{figure}{0}
\renewcommand{\thefigure}{S\arabic{figure}}
\part*{Supplementary Information}
\section*{Sample quality}
\begin{figure}[h]
\includegraphics[keepaspectratio,clip,width=0.8\textwidth]{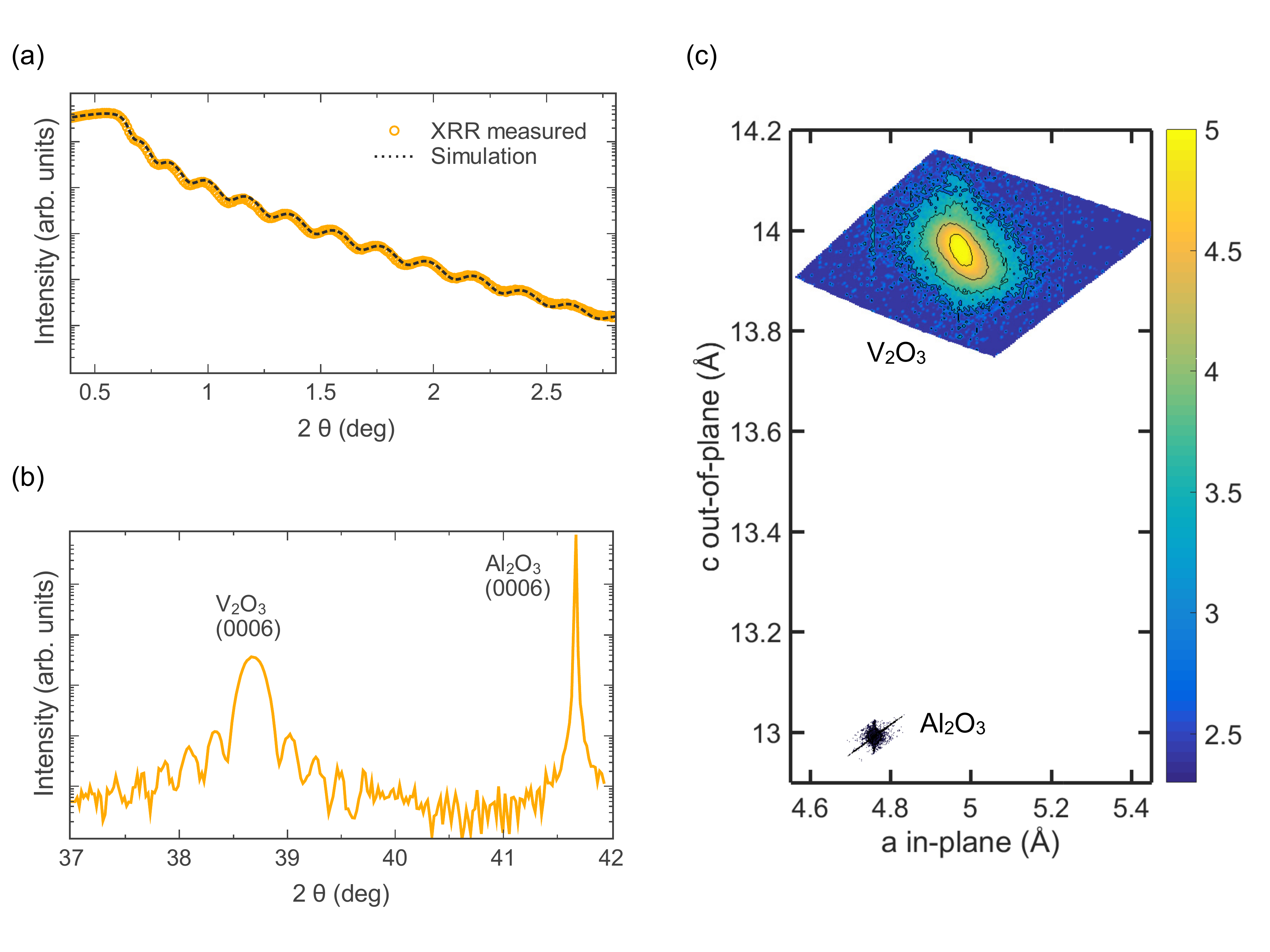}
\caption{(a) XRR measurement and simulated curve to extract the film thickness and surface roughness. (b) XRD $\theta$-2$\theta$ scans around the (0006) reflections of the 40 nm V$_2$O$_3$ film and Al$_2$O$_3$ substrate. (b) RSM transformed to in-plane and out-of-plane lattice spacings of the film and the substrate.} 
\label{Fig_S1}
\end{figure}

To address the structural quality of the 40 nm V$_2$O$_3$ film epitaxially grown on (0001)-Al$_2$O$_3$, X-ray diffraction measurements were performed on a Panalytical X'pert Pro diffractometer using monochromatic K$\alpha_1$ radiation of a Cu anode. 
A film thickness of 40 nm and a surface roughness below 1 nm were determined from X-ray Reflectivity (XRR) measurements and the simulation of the data with the X'pert Reflectivity software. The XRR measurement and simulated curve are shown in Fig. \ref{Fig_S1}a.

The $\theta$-2$\theta$ scan in Fig. \ref{Fig_S1}b shows the symmetric peak (0006) of the film and substrate, thus confirming the single crystalline nature of the film, with the $c$-axis oriented perpendicular to the surface. The finite-size oscillations (Pendell\"osung fringes) around the film peak maximum confirm the high quality epitaxial deposition with a smooth surface and interface.

Finally, to extract the lattice parameters ($a$ and $c$) of the film, Reciprocal space maps (RSM) around the asymmetric peak (1 0 -1 10) of the film and the substrate were performed. The resulting RSM transformed to in-plane ($a$-axis) and out-of-plane ($c$-axis) lattice spacings is shown in Fig. \ref{Fig_S1}c. The lattice parameters extracted from the peak maximum are $a$=4.977 {\AA} and $c$=13.962 {\AA} and correspond well with the bulk V$_2$O$_3$ values of $a$= 4.954 {\AA} and $c$=14.008 {\AA}. The small tensile (compressive) in-plane (out-of-plane) strain of 0.46\% (-0.33\%) is thermal in nature \cite{Dillemans2014}.

\section*{PEEM resolution}
\label{section_S2}
\begin{figure}[t]
\includegraphics[keepaspectratio,clip,width=1\textwidth]{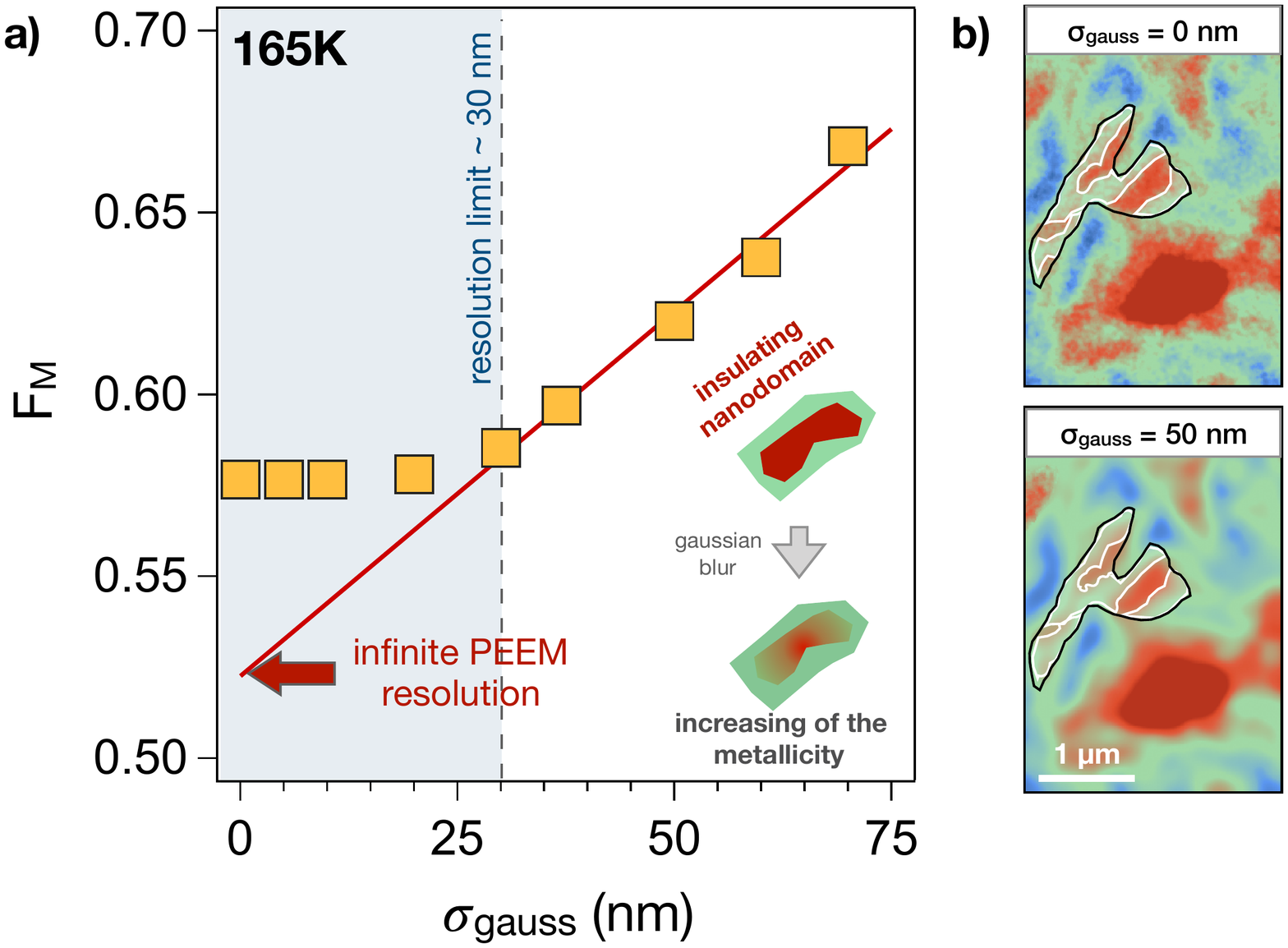}
\caption{a) Estimated metallic filling fraction $F_M$ (yellow markers) as a function of radius of the gaussian blurring of the  XLD-PEEM images. Above the instrumental resolution limit ($\sim$30 nm) a linear increase of the estimated $F_M$ is observed. The linear fit (red line) allows us to extrapolate the real $F_M$ that would be observed by an infinite resolution experiment. The difference between the extrapolated and the calculated metallic fraction can be considered as the uncertainty of the points of heating branch of the hysteresis cycle shown in Fig. \ref{Fig_dynamics}. b) Top: XLD-PEEM images taken at 165 K. The metallic regions are highlighted by the white contours. Bottom: image blurred with $\sigma_{\mathrm{gauss}}$=50 nm. After the pixel-based segmentation, the metallic area (white shadows) increases by about 4\%.} 
\label{Fig_S2}
\end{figure}

The structures observed in XLD-PEEM images are the result of a convolution between the real size of the spatial domains and the experimental resolution ($\sim$30 nm). The finite resolution prevents the identification of insulating areas with dimension less than 30 nm, thus leading to an over-estimation of the real metallic filling fraction $F_M$ determined via the clustering process of the pixel-based segmentation routine. In order to quantify this over-estimation, we simulated the effect of the experimental resolution by performing a gaussian blurring in the same region of interest of Fig. \ref{Fig_dynamics}a. The plot in Fig. \ref{Fig_S2} reports the estimated $F_M$ as a function of the radius ($\sigma_{\mathrm{gauss}}$) of the gaussian blurring. A linear fit has been used to extrapolate the value of $F_M$ in the $\sigma_{\mathrm{gauss}}\rightarrow$ 0 limit (ideal case) and therefore the uncertainty in the metallic filling fraction $F_M$, which amounts for instance to $\sim$10\% at 165K.

\section*{reflectivity variation and Bruggeman Effective Medium Theory}
\label{section_S3}
The Bruggeman effective medium approximation has been widely used (see e.g. Refs. \onlinecite{Bruggeman1935,Jepsen2006,Lupi2010}) to describe the effective optical properties of vanadium oxides across the percolative $IMT$. The model relates the effective macroscopic dielectric function ($\epsilon_{eff}$) of an inhomogeneous system, constituted by a mixture of metallic and insulating regions, to the dielectric functions of the metallic ($\epsilon_{M}$) and insulating ($\epsilon_{I}$) bulk species:  
\begin{equation}
\epsilon_{eff}=\frac{1}{4}\{\epsilon_{I}(2-3F_{M})+\epsilon_{M}(3F_{M}-1)+\sqrt{\left[\epsilon_{I}(2-3F_{M})+\epsilon_{M}(3F_{M}-1)\right]^2+8\epsilon_{I}\epsilon_{M}}\}
\label{eq_BGT}
\end{equation}
where $F_{M}$ is the metallic filling fraction as defined in the main text.
The procedure adopted for calculating the effective optical properties is the following: 
\begin{enumerate}
\item $\epsilon_{M}$ is extracted from the experimental high-temperature temperature ($T$=200 K) reflectivity (see the inset of Fig. S3a) by  means of a multi-reflection model, which accounts for the thin V$_2$O$_3$ film and the substrate. The input dielectric function, which perfectly reproduces the measured reflectivity in the 1.4-2.2 eV energy range ($R_M(\omega)$), is given by a Drude-Lorentz model (DLM). The oscillators parameters of the DLM have been taken from Ref. \citenum{Stewart2012}; 

\item the experimental reflectivity change between the metallic and insulating phases (shown in Fig. \ref{Fig_optics} of the main text) in the spectral region 1.4-2.2 eV is reproduced by modifying the central frequency of the Lorentz oscillator centered at 2 eV ($e^{\pi}_g$ LHB$\rightarrow$UHB transition). As output, we obtain $\epsilon_{I}$, which reproduces $R_M(\omega)$-$R_I(\omega)$, where $R_I(\omega)$ is the thin film reflectivity at $T$=120 K;

\item the effective dielectric function, $\epsilon_{eff} (F_M,\omega)$, is calculated by means of Eq. \ref{eq_BGT} for the $F_M$ values extracted from the PEEM images (Fig. \ref{Fig_dynamics});

\item the non-equilibrium variation of the effective dielectric function is calculated as $\delta \epsilon_{eff} (F_M,\omega)$=$\epsilon_{eff} (F_M+\delta F_M,\omega)$-$\epsilon_{eff} (F_M,\omega)$;

\item the relative reflectivity variation is calculated as $\frac{\delta R}{R}(F_M,\omega)$=[$R(F_M+\delta F_M,\omega)$-$R(F_M,\omega)$]/$R(F_M,\omega)$ using the multi-reflection model and $\epsilon_{eff}(F_M,\omega)$ as an input.
\end{enumerate}

By using the above procedure, we numerically demonstrate the possibility of linearizing both $\delta \epsilon_{eff}(F_M,\omega)$ and $\frac{\delta R}{R}(F_M,\omega)$. In Fig. S3a we plot the real and imaginary parts of the calculated $\delta \epsilon_{eff} (F_M,\omega)$ for $F_M$=0.5 and $\delta F_M$=0.1. The $\delta \epsilon_{eff} (F_M,\omega)$ function perfectly overlaps to $[\epsilon_{M}(\omega)-\epsilon_{I}(\omega)]\delta F_{M}$. 
Similar results are obtained for any value of $F_M$. 

In Fig. S3b we numerically demonstrate the relation $\Phi(F_M,\omega)\simeq [R_M(\omega)$-$R_I(\omega)]$/$R_I(\omega)]$, which is used in this work to extract the transient filling factor variation from the relative reflectivity variation. The function $\Phi(F_M,\omega)$ is calculated as:
\begin{equation}
\Phi(F_M,\omega)=\frac{\delta R}{R}(F_M,\omega)\frac{1}{\delta F_M}
\end{equation}
for $F_M$ ranging from 0.01 to 0.93. $\Phi(F_M,\omega)$ is almost independent of the initial value of $F_M$ and, within the error bars, is equal to the experimental $[R_M(\omega)$-$R_I(\omega)]$/$R_I(\omega)]$ function, where $R_I(\omega)$ is the reflectivity at $T$=120 K. In the inset of Fig. 9S3b we also plot the calculated $\frac{\delta R}{R}(\hbar\omega$=2 eV) as a function of $\delta F_M$, thus proving the linear relation between the pump-induced reflectivity variation in the visible range and the filling fraction variation.

Finally, in order to demonstrate that the linearization $\delta R(F_M,\omega)/R$=$\Phi(F_M,\omega) \cdot \delta F_M$ holds in the entire fluence range spanned by our experiment (see the fluence-dependent data in Fig. \ref{Fig_avrami}c), we calculated $\delta R(F_M,\omega)/R$ at the filling factor corresponding to $T$=145 K, i.e. $F_M \simeq$0.01, for $\delta F_M$ extending up to 30\%. The estimated reflectivity variation is linearly proportional to $\delta F_M$ for the values accessed by the experiment, as shown in the inset of Fig. S3b. Similar results are obtained at any value of $F_M$.

\begin{figure}[h]
\includegraphics[keepaspectratio,clip,width=1\textwidth]{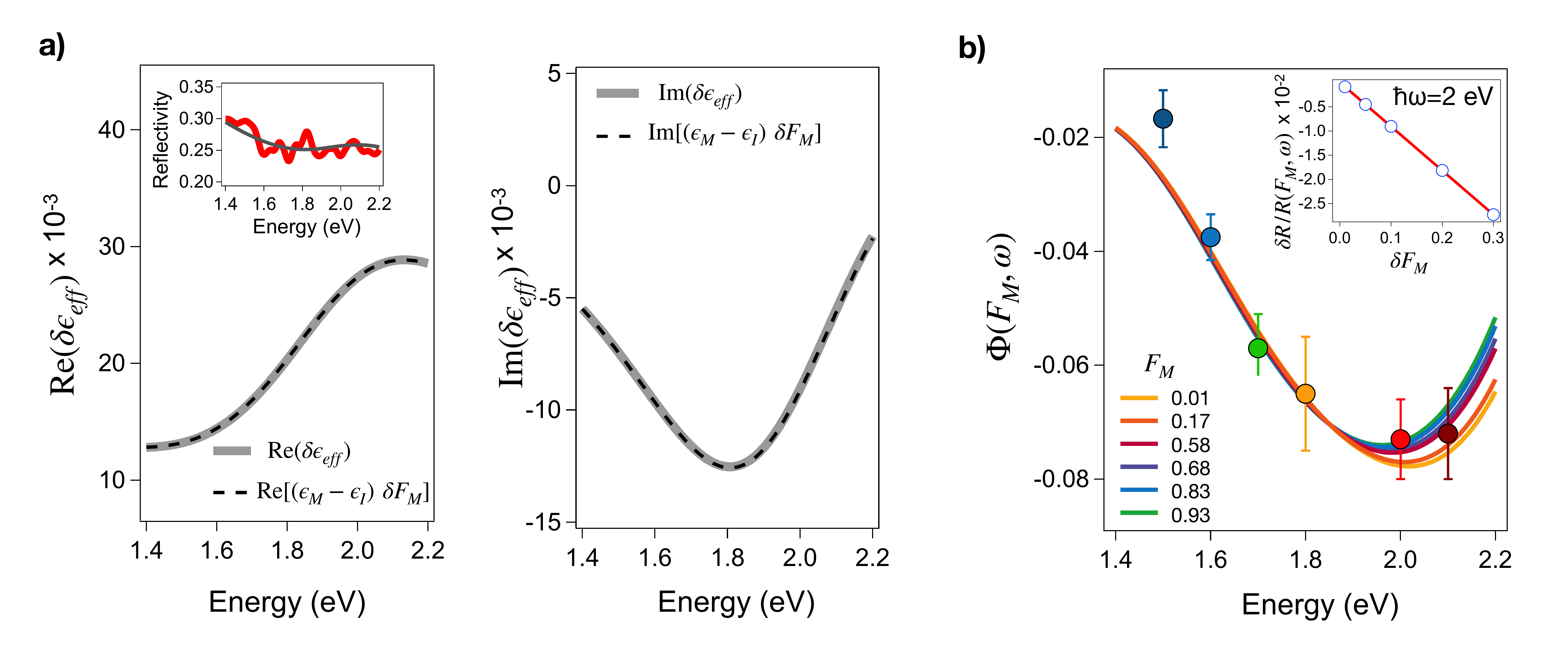}
\label{Fig_supp3}
\caption{\textbf{Effective dielectric function model}: a) Frequency dependence of the real and imaginary part of the out-equilibrium effective dielectric function $\delta\epsilon_{eff}(F_M,\omega)$ calculated by means of the Bruggeman model (gray line) for $F_{M}=0.5$ and $\delta F_{M}=0.1$. The black dashed line represents the approximation  $\delta\epsilon_{eff}\simeq(\epsilon_{M}-\epsilon_{I})\delta F_{M}$. The two curves differ by less than 0.2 \% at 2 eV photon energy. Inset: high-temperature ($T$=200 K) reflectivity as a function of the photon energy (red line) and Drude-Lorentz reflectivity model for a 40 nm V$_{2}$O$_{3}$ film (black line) on a Al$_2$O$_3$ substrate. b) $\Phi(F_{M},\omega)$=[$R_{M}(F_{M},\omega)$-$R_{I}(F_{M},\omega)]/R_{I}(F_{M},\omega)$ plotted for different values of $F_{M}$, corresponding to the metallic filling fractions at the temperature reported in Fig. \ref{Fig_optics}. The function $\Phi(F_{M},\omega)$ is almost independent of $F_M$ and differs by less than 5\% in the energy range 1.8-2.2 eV an for 0.01$<F_M<$0.93. The colored dots represent the equilibrium reflectivity change, $\left[R_{M}(\omega)-R_I(\omega)\right]/R_I(\omega)$, as shown in Fig. \ref{Fig_optics}. Inset: Linear relation between $\delta R/R (\hbar\omega$=2 eV) for $F_M$=0.01 (dots) and $\delta F_{M}$. The red thin line is the linear fit to the data.} 
\end{figure} 

\end{document}